\documentclass[12pt,preprint]{aastex}
\usepackage{epsfig}
\usepackage{amsmath,amsfonts,amssymb}
\usepackage{subfigure}
\newcommand{\sqarcsec}{\ensuremath{\Box^{\prime\prime}}}
\newcommand{\sqarcdeg}{\ensuremath{\Box^{\circ}}}

\newcommand{\ctbd}[1]{}

\slugcomment{To appear August 2005 issue of PASP}

\shorttitle{The XO Project: I. Equipment}
\shortauthors{McCullough et al.}

\begin{document}

\title{The XO Project: Searching for Transiting Extra-solar Planet Candidates}

\author{P.~R.~McCullough\altaffilmark{1,2}
J.~E.~Stys\altaffilmark{1}, 
J.~A.~Valenti\altaffilmark{1},   
S.~W.~Fleming\altaffilmark{3},  
K.~A.~Janes\altaffilmark{4} and
J.~N.~Heasley\altaffilmark{5}
}
\email{pmcc,jstys,valenti@stsci.edu; scfleming@vassar.edu; janes@bu.edu; heasley@ifa.hawaii.edu}

\altaffiltext{1}{Space Telescope Science Institute, 3700 San Martin Dr., Baltimore MD 21218}
\altaffiltext{2}{University of Illinois, Urbana, IL 61801}
\altaffiltext{3}{Vassar College, Dept. of Physics and Astronomy, 124 Raymond Ave., Poughkeepsie, NY 12604-0745}
\altaffiltext{4}{Boston University, Astronomy Dept., 725 Commonwealth Ave.,
Boston, MA 02215}
\altaffiltext{5}{University of Hawaii, Inst. for Astronomy, 2680 Woodlawn Dr., Honolulu, HI 96822-1839}

%% ========================== ABSTRACT ==============================
\begin{abstract}

The XO project's first objective is to find hot Jupiters transiting bright
stars, i.e. V $<~12$, by precision differential photometry.
Two XO cameras have been operating
since September 2003 on the 10,000-foot Haleakala summit on Maui. Each
XO camera consists of a 200-mm f/1.8 lens coupled to a 1024x1024 pixel,
thinned CCD operated by drift scanning.
In its first year of routine operation, XO has observed 6.6\% of the
sky, within
six 7\arcdeg-wide strips scanned from 0\arcdeg\ to $+63$\arcdeg\ of
declination and centered at RA=0, 4, 8, 12, 16, and 20 hours. 
Autonomously
operating, XO records 1 billion pixels per clear night, calibrates
them photometrically and astrometrically, performs aperture photometry,
archives the pixel data and transmits the photometric data to STScI for
further analysis. From the first year of operation, the resulting database
consists of photometry of $\sim$100,000 stars at more than 1000 epochs
per star with differential photometric precision better than 1\%
per epoch. Analysis of the light curves of those stars produces transiting-planet
candidates requiring detailed follow up, described elsewhere, culminating in
spectroscopy
to measure radial-velocity variation in order to differentiate genuine
planets from the more numerous impostors, primarily eclipsing binary and multiple
stars.
\end{abstract}

%% ========================== KEYWORDS ==============================

% ========================= INTRODUCTION =================================
\section{Introduction}
\label{sec:intro}

Borucki \& Summers (1984) proposed detection of planets with the transit
technique. At the time, their proposal was thought to be impractical
because astronomers expected planetary systems to be like our own. In
particular the Jovian-sized planets that could create a readily-detectable
photometric transit would occupy orbits many AU in radius with periods of
many years.  Since the discovery of the ``hot Jupiter'' orbiting 51 Peg
with a 4.23 day orbit, many attempts have been made to find hot Jupiters
that transit stars bright enough ($m_V \la 12$) for detailed studies to be
performed with existing telescopes such as HST, Keck and the VLT (e.g. Vulcan,
Stare, etc).  The first such success was Tres-1 (Alonso et al. 2004).
The OGLE collaboration's successes were with fainter stars (Udalski et al. 2003).

The radius of solar-composition objects
is expected to change by less than a factor of two (e.g. Burrows et al. 2001)
over the entire range in mass from the bottom of the main sequence
to less than a Jupiter mass.
So transit observations by themselves provide no information
about the mass of the planet (or brown dwarf, or red dwarf).  On the other
hand, radial velocity studies yield only minimum masses. However,
the combination of the Doppler orbit and the observation of transits yields
the gross physical characteristics of the planet: mass, radius, density, and
``surface'' gravity.
A transiting planet is interesting in at least two other ways: 
1) absorption of starlight is a much
larger signal than reflected starlight, so for example absorption
spectroscopy has already permitted detection of an exoplanet's
atmosphere (Charbonneau et al. 2002), and 2) the rapid and precisely
predictable on/off nature of the transit permits excellent calibration,
which among other things, allows one to search for natural satellites and
Saturn-like rings orbiting the transiting planet and to attempt to
measure the planet's albedo by observing the reflected light of the
planet being blocked by the star (Brown et al. 2001).

This paper describes the XO project's design, implementation,
and verification.  Not an acronym but a name, XO is pronounced as
it is in ``exoplanet.''  We describe the XO design requirements in
Sec. \ref{sec:rqmt}, the hardware in Section \ref{sec:hardware}, the
observing strategy in Section \ref{sec:observing}, and the software in
Section \ref{sec:software}.  Section \ref{sec:ver} shows that
our system is finding transiting hot Jupiter (THJ) candidates. 
In future papers we describe
additional observations that test whether a candidate is definitely an
impostor such as an eclipsing binary star, or potentially one of the
THJs that we seek (McCullough et al. 2005).

% ========================== REQUIREMENTS ================================
\section{Requirements}
\label{sec:rqmt}

Pepper et al. (2003) formalize the optimization of systems designed to
find THJs and have selected a 2-inch diameter telescope with a 4kx4k
sensor as best for a $2\pi$ steradian survey of stars with V$\la 10$.
Independently we designed the XO system with a similar but simpler analysis
for a limiting magnitude of V$\la 12$, requiring a larger diameter aperture,
d $\approx 0.1$ m, and in order to accommodate drift scanning,
a smaller instantaneous field of view, 7\arcdeg$\times$7\arcdeg .

The XO system was designed to find THJs around stars
bright enough to permit significant follow up as described in the introduction,
i.e. ($m_V \la 12$).
The number of THJ-systems on the sky is estimated below
from the number density of stars as a function of their brightness,
the frequency of hot Jupiters around those stars,
and the geometric probability of the orbit being inclined such that a transit
can occur as seen from Earth.
The photometric precision required is set by the fraction of the
star's area obscured by the THJ, $\sim 0.01$. The cadence of the observations
is set by the need to have multiple observations made during the duration
of the transit, $\sim 0.1$ day. The duration of an observing sequence is set
by the need to observe multiple transits to define a tentative ephemeris. Given that
useful observations are obtained only at moderate zenith angles during clear nights, the observing
sequence must be much longer than the orbital period in order to witness at least 3 transits
(Brown 2003).

Of stars brighter than $\rm m_V = 12$ at the north galactic pole (NGP), 
there are 2.9 main sequence stars per \sqarcdeg\ with $\rm M_V =$ 4.5 to 5.5,
and 1.3 with $\rm M_V =$ 5.5 to 6.5 (Bahcall \& Soneira 1981).
Solar-type and later stars with $\rm M_V > 5$\ and brighter 
than our limiting magnitude of $\rm m_V = 12$ must be closer than $D = 250$ 
pc, whereas for such stars the scale height $H = 400$ pc. For a
volume density of stars with exponential scale height H, the
ratio of stars per square degree within a distance D at
Galactic latitude $b=0$\arcdeg\ to those at $b=90$\arcdeg\ is
$D/(H\times(1-exp(-D/H))$, which equals 1.34 for D/H = 250 pc / 400 pc = 0.625,
and approximately equals 1 + 0.58*D/H for D/H $\le 1$.
The density enhancement in the galactic plane is countered by the
disadvantages of
crowding and confusion. For XO's drift-scanned observations of a variety of
Galactic latitudes, we empirically find the maximum surface density of stars
with photometry sufficient for our purposes occurs at $b \approx\ 20$\arcdeg (Section \ref{sec:ver}).
Although stars become more concentrated
toward the Galactic plane with earlier type, early-type stars are larger in
physical size and thus are expected to have transits of lower amplitude that
are more difficult to detect.
For simplicity, we use the stellar density at the NGP in order to conservatively estimate
that there are $\sim$160,000 solar type stars with $\rm m_V <
12$ over the entire celestial sphere. 
The probability that a given solar type star has a THJ is 0.00075,
because the fraction of stars with
Jovian planets is 0.05 (Marcy \& Butler 2000), the fraction of those
planets with periods less than 7 days is 0.15 (Brown 2003), and the
fraction of those that could exhibit transits seen from Earth is 0.10
(Borucki \& Summers 1984).  
Thus, we expect 120 THJs orbiting solar type
stars brighter than 12th magnitude, or 30 THJs orbiting stars brighter
than 11th magnitude. The corresponding predictions are 7.5, 2, and 0.5
stars for $\rm m_V < 10,~9,~and~8$, respectively.

There are at least three interesting points implied by the calculations above:
\begin{enumerate}
\item{the 8th magnitude HD 209458 probably is the brightest star exhibiting
hot-Jupiter transits,}
\item{there is likely one hot Jupiter
transiting an 11th magnitude star for every 1400 \sqarcdeg\ of sky, and}
\item{at any given time approximately one bright ($\rm m_V < 11$) star is being transited.}
\end{enumerate}

For simplicity, we have assumed the
probabilities are independent and can be multiplied together. A thorough
analysis of joint probability functions is beyond the scope of this
paper, and indeed some of the joint probabilities are unknown (Brown 2003). 
Pepper et al. (2003) estimate there are $\sim$5 THJs orbiting
stars brighter than V=10 mag for $\rm 4.5 < M_V < 5.5$, and scaling from their
Figure 1 we estimate for V$<$10 an additional $\sim$6 THJs and $\sim$3 THJ for
$\rm 3.5 < M_V < 4.5$ and $\rm 5.5 < M_V < 6.5$, respectively,
if the frequency of THJs is independent of $\rm M_V$.
We have based our requirements upon the statistics from radial velocity
surveys, which are secure for solar type stars. However, if the estimate
of Pepper et al. (2003) is appropriate, then we could hope for
of order 200 THJs orbiting approximately solar sized (or smaller) stars
brighter than 12th magnitude.

Succinctly, the XO system is required to image hundreds of square degrees of sky
many times per hour for months, and from those images the software must enable
photometry of stars $9 \la m_V \la 12$ with a
precision of $\sim 10$ millimag per measurement. Those requirements have been met by 
the XO Mark I system described below. Given the demonstrated performance of the Mark I
system, we anticipate replicating it to speed the discovery of THJs.

\section{Hardware Implementation}
\label{sec:hardware}

\begin{figure}[!ht]
\begin{center}
\epsscale{0.8}
\plotone{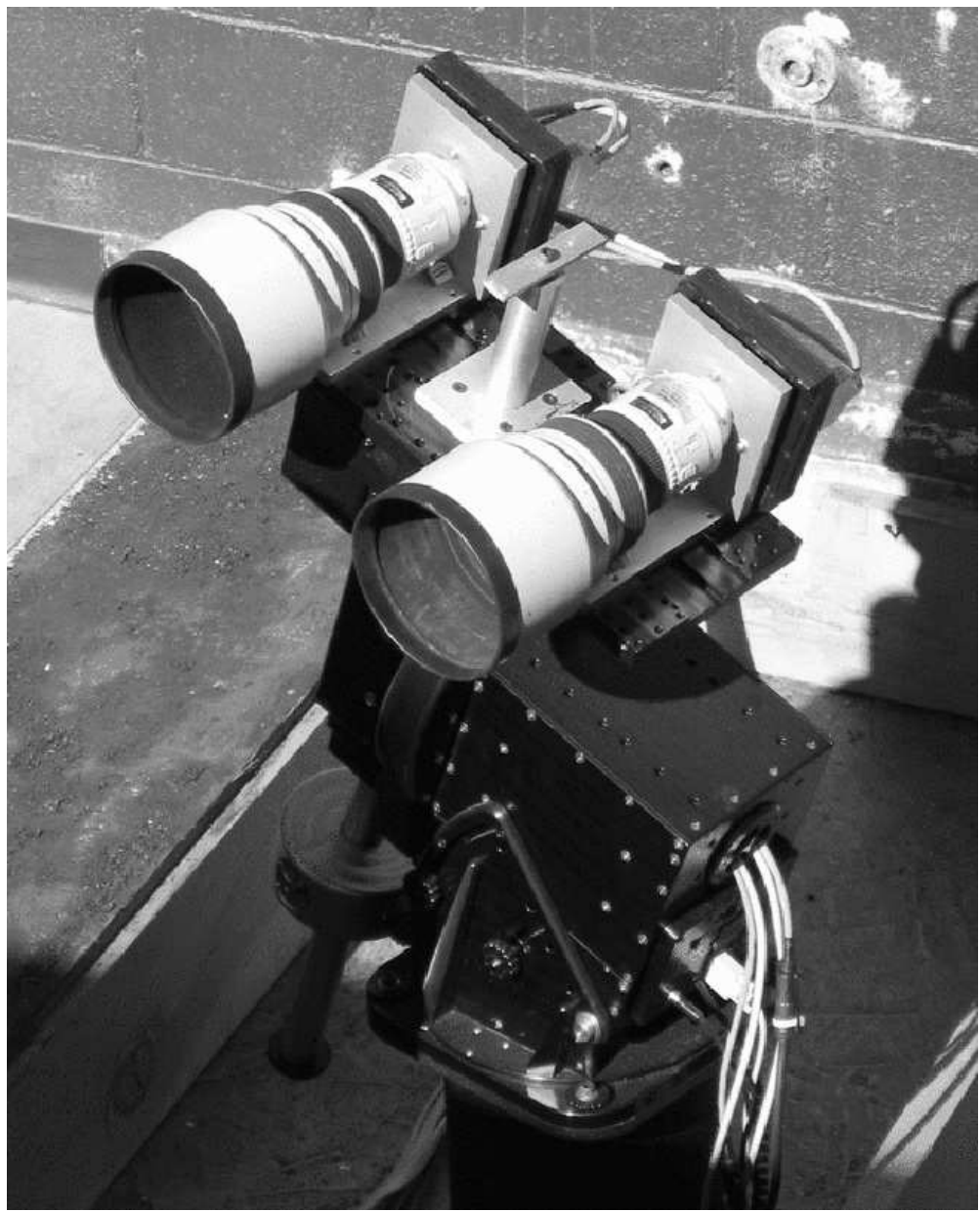}
\caption{The XO Mark I observatory: two 200\,mm f/1.8
lenses and 1K$\times$1K CCDs attached to a German equatorial mount,
deployed under a roll-off roof.
\label{fig:cameras}
}
\end{center}
\end{figure}

\begin{figure}[!ht]
\begin{center}
\epsscale{0.8}
\plotone{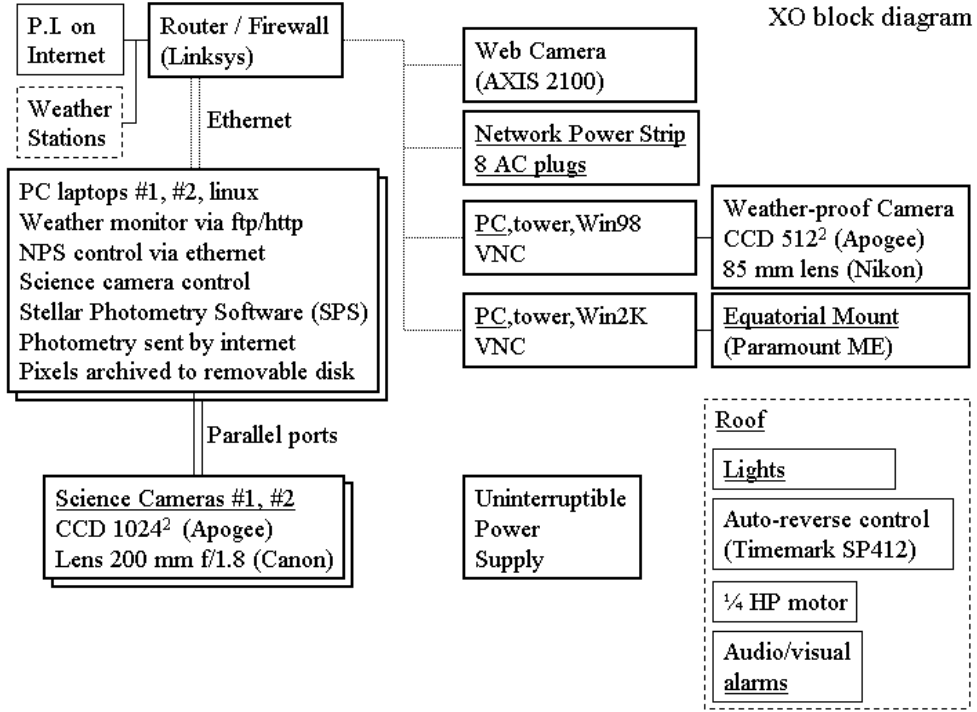}
\caption{The XO Mark I system block diagram.
\label{fig:blockdiagram}
}
\end{center}
\end{figure}

The XO Mark I hardware is described in this section. 
Figure \ref{fig:cameras} illustrates the cameras and equatorial mount;
Figure \ref{fig:blockdiagram} gives a block diagram of the system;
and
Table \ref{tbl-1} summarizes the components.

Many systems have been built to discover THJs orbiting bright stars
(Horne 2003).
Some of the unique aspects of the XO system are that it uses
a broad spectral bandpass (0.4 $\mu$m to 0.7 $\mu$m) to collect more
photons per second, drift scanning to simplify calibration, 
aperture photometry for simplicity and reliability,
and two identical cameras pointed in the same direction to collect more
photons and to provide redundancy so that observing will not be entirely
interrupted due to failure of a single camera or its control computer.

Sometimes a component is unresponsive and must be reset by cycling its AC power.
We accomplish this either by human intervention either in person or
remotely using a network power strip that permits remote control
of 8 AC power outlets independently (Figure \ref{fig:blockdiagram}).
The network power strip is one of the essential components in the XO system
and it has operated 100\% reliably. 

\begin{table}
\caption{Summary of the XO Mark I Equipment\label{tbl-1}}
\footnotesize
\vspace{3mm}
\begin{tabular}{|l|ccc|}
\tableline\tableline
{\it Location:}&&&\\
Haleakala, Maui&{\it  Longitude}:&{\it Latitude}:&{\it Elevation}:\\
&$156^{\circ}$15'.3W&$20^{\circ}$42'.4N &3054 m\\ \hline
{\it Shelter:}&&&\\
roll-off roof&Cable-&1/4 HP gear motor&{\it Surveillance camera}\\
&driven&Grainger Inc. 5K942&AXIS Inc. 2100\\ \hline
{\it Mount:}&&&\\
Paramount ME&German-&{\it Scan rate in RA:}&{\it Scan rate in Dec.:}\\
Software Bisque Inc.&Equatorial&sidereal&478 arcsec$\cdot$sec$^{-1}$\\ \hline
{\it Objective:}&&&\\
Canon EF200 lenses (two)&{\it Diameter:}& {\it Focal ratio:}& {\it Plate scale:}\\
& 0.11 m & f/1.8  & 1.058 ''/$\mu$m\\ \hline
{\it Filter:}&&&\\
50 mm diameter, 3.3 mm thick&{\it Bandpass:}&{\it Transmission:}&${\lambda}_{cutoff}$:\\
Edmund Optics Inc. NT54-517&flat over&$\sim$ 95\%&700 nm\\
&$400 nm < \lambda < {\lambda}_{cutoff}$&&\\ \hline
{\it Detector:}&&&\\
Apogee Ap8p CCD& 1024 x 1024&{\it FOV:}&{\it Pixel scale:}\\
thinned&24$\mu$m pixels&$7^{\circ}.2\times7^{\circ}.2$&25.4 ``/pixel\\ \hline
\tableline\tableline
\end{tabular}
\end{table}

%----------------
%-----------

%----------------
%-----------

%----------------
\begin{figure}[!ht]
\begin{center}
\epsscale{1.0}
\plotone{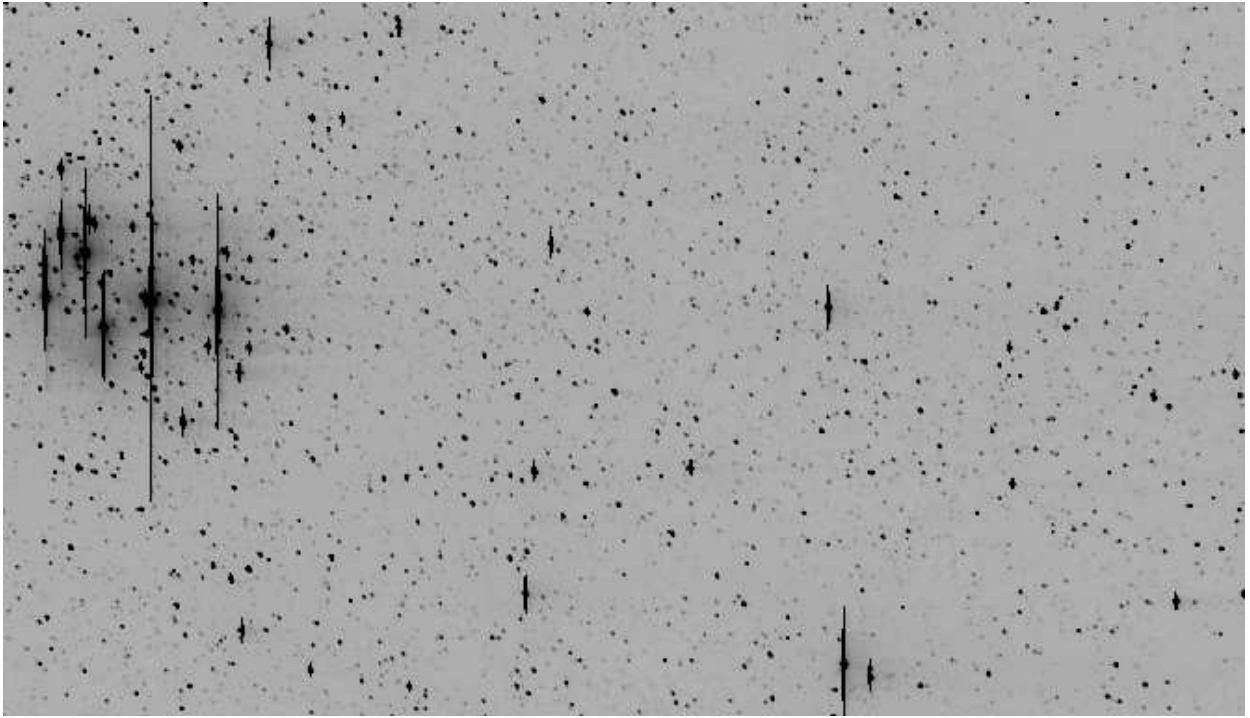}
\caption{A representative section
of a flat-fielded XO image shows the Pleiades star cluster (left side).
This 1024-column section is as wide as the XO images, 7.2\arcdeg\ but is
only 6\% of the 9000-row height of the scans.
The vertical and horizontal trails from very saturated stars do not prevent good
photometry of most of the stars in each image.
\label{fig:image}
}
\end{center}
\end{figure}
%-----------

%----------------
\begin{figure}[!ht]
\begin{center}
\epsscale{1.0}
\plotone{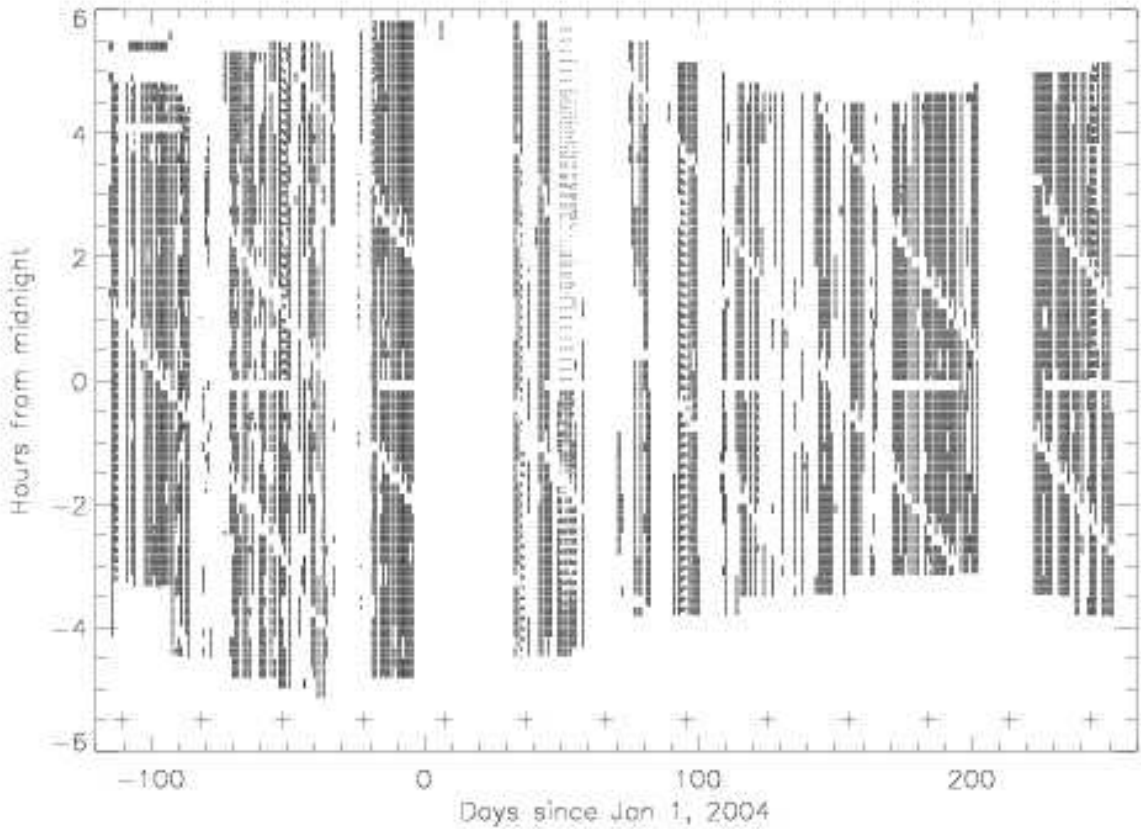}
\caption{The periods of time spent observing during the first year of operation
are plotted. The dates of full moon are indicated with plus signs.
\label{fig:year1}
}
\end{center}
\end{figure}
%-----------

%----------------
\begin{figure}[!ht]
\begin{center}
\epsscale{1.0}
\plotone{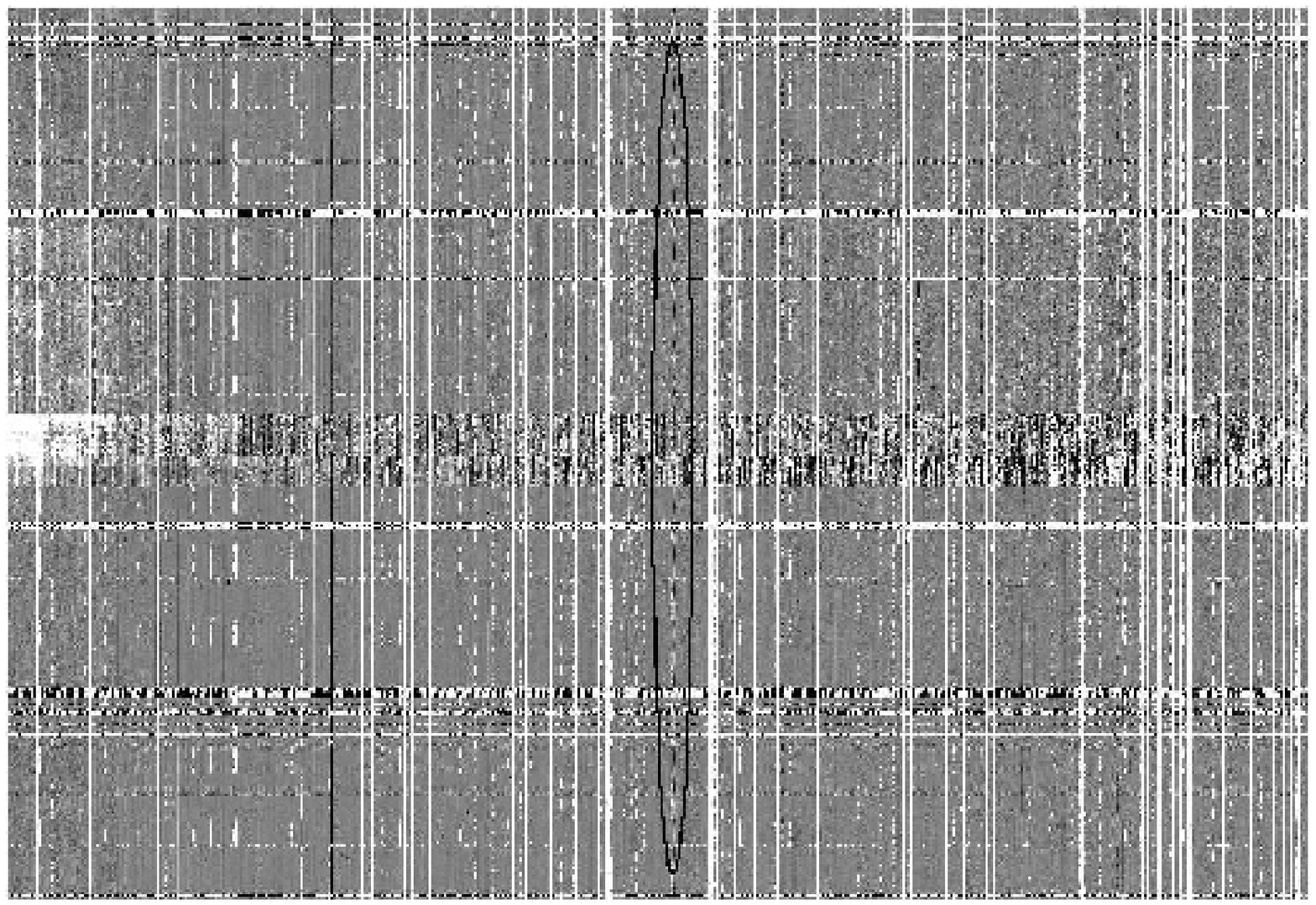}
\caption{This figure shows a representative section of a calibrated
array of differential stellar magnitudes, with each column corresponding
to a single star, and each row to a single epoch. The mean magnitude
of each star has been subtracted from each column. Bad data have not been flagged.
Ideally the entire
image would be white noise, less at the left (brighter stars) and more
at the right (fainter stars). In practice data are missing (white) and
trends are evident (see text). Visual inspection of diagrams like this
reveals trends and assists in improving the calibration. Large amplitude
variable stars are also apparent; an example near the middle of the figure
is shown inside the ellipse.
\label{fig:starvsepoch}
}
\end{center}
\end{figure}

% ========================== ROOF CONTROL ================================
\subsection{Weather Sensing and Roof Control}
\label{subsec:roof}

Gaustad et al. (2001) enlisted a human operator of an adjacent facility
to authorize or to override, via email, the nightly opening and closing
of the dome protecting their robotic telescope.  The XO system observes
without human assistance. Weather hazardous to the equipment is sensed
in near real time as described below, and weather that is not hazardous
but is unsuitable for observing is sensed after the fact by the science
data analysis software.

We use three sources of weather data to determine whether to open our
roof. The first is a CCD camera to detect stars. The second source is a
set of weather stations operated by other tenants on the mountain. The
third is a forecasting service operated by yet another tenant.  If all
three services are operating, then the data from every one must meet
specific criteria for our system to open its roof. If one of the second
or third services is inoperable, we reconfigure our system to ignore it.
These three sources of weather data are described in turn below.

A small camera mounted in a weather-proof enclosure is aimed near the
north celestial pole. At regular intervals, the PC takes a short exposure
of a $\sim 3$-degree field of view.
If the sky is clear, many stars are visible in the image and our software
can determine an astrometric solution in the same manner as it does for
our science images (Section \ref{subsec:datared}). However, with time
the images became increasingly defocused, due to wind shaking loose the
lens' helical focus mechanism. 
To compensated for this issue remotely,
we adopted a different and more robust algorithm that works
well with even very defocused images.  The algorithm operates on two
images taken twenty minutes apart. It rotates one image about the pole
in order to register it to the other image, and if the two-dimensional
cross-correlation function peaks within $\pm$3 pixels of the center
and is greater than an empirically-determined threshold (0.17), then we
assume the sky is clear. The latter algorithm is simple and effective,
so we have not returned to the original algorithm of matching patterns of
stars, even after re-adjusting the focus.
We note that pointing the camera near the pole has some advantages:
1) sunlight cannot reach the lens and thereby damage a shutter, 2) rain
(or snow) slides off the inclined window, and 3) clouds tend to come
from up from below on Haleakala, so we think aiming our weather camera
at a large zenith angle gives early warning of fog rising up the mountain.

If the data from any of the weather stations triggers any of the following conditions, then we close the
roof, or keep it closed: 
1) data are non-existent or stale, i.e. older than 30 minutes,
2) humidity $>$ 75\% or dew point $< 4$\arcdeg~C from ambient,
3) temperature $< -5$\arcdeg~C,
4) wind velocity $> 20$ m s$^{-1}$,
5) non-zero rain accumulation in the past 10 minutes,
6) sunlight detected on a solar cell.
Humidity and/or dew point accounts for nearly all closures.
Because the humidity on Haleakala tends to be quite bimodal, either very low or nearly 100\%,
and because high humidity should correlate with poor photometric precision,
we have not attempted to optimize the thresholds for humidity and dew point.

Another tenant on Haleakala operates weather sensors and a neural network
prediction of the probability of inclement weather in the near future
(a ``forecast'') and now (a ``nowcast'').  XO closes its roof if the
``forecast'' or ``nowcast'' indicates a 95\% or greater probability of
inclement weather.

% ========================== DRIFT SCANNING ==============================
\subsection{Drift Scanning}
\label{subsec:ds}

\begin{figure}[!ht]
\begin{center}
\epsscale{1.0}
\plotone{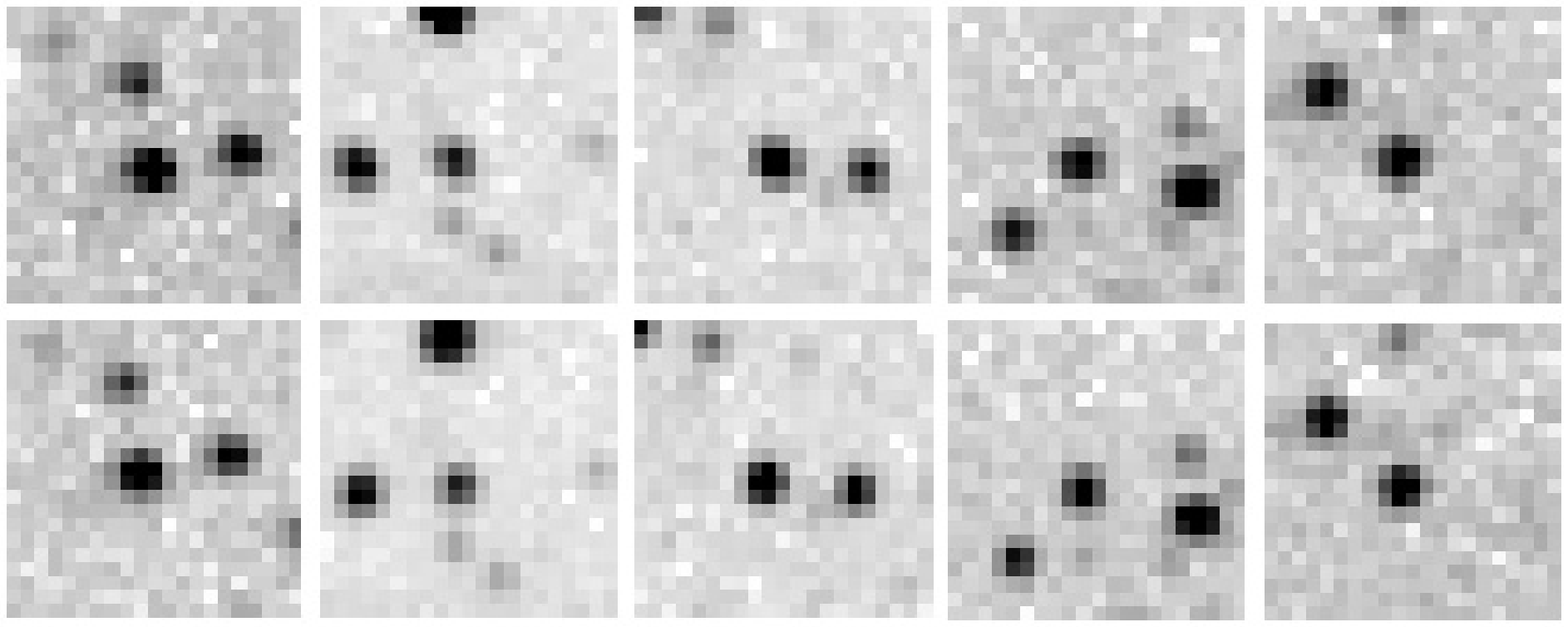}
\caption{Images of stars are nearly uniform across the field of view.
Subimages centered at CCD columns 100, 300, 500, 700, and 900 (left to right) 
are displayed for camera 0 (top) and camera 1 (bottom). The two cameras observe
nearly the same field of view simultaneously; in their native format,
they are misaligned by 13 columns, i.e. 1.3\% of
the sensor's width, but we have shifted one by 13.0 columns for this plot.
\label{fig:psf}
}
\end{center}
\end{figure}

Drift scanning is efficient if the number of rows readout is much larger
than the number of columns, i.e. for long, rectangular fields of view.
Its advantage over staring-mode imaging is that the flat-field and the
dark correction are homogenized by shifting the charge through all the
rows. Thus, the drift-scanned vectors are more uniform and smoother than
their staring-mode counter parts, which are 2-D arrays.
The drift-scanned PSF is slightly wider in both directions compared to what
it would be with ``standard'' staring-mode observations, but this is
beneficial in the following sense.
Drift scanning makes
intra-pixel gain variations irrelevant; even intra-column variations are
homogenized due to the fact that stars do not track perfectly parallel
to columns. 
The drift-scanned PSF is approximately a Gaussian with FWHM = 1.8 pixels.
During the first year of operation, the correlation coefficient of the
observed PSF with a FWHM=1.8 pixel Gaussian is $\sim0.80$ with no measurable
dependence on time of night, day of year, hour angle, or
temperature from 3\arcdeg\ C to 16\arcdeg\ C. The correlation coefficient
is expected to be variable with position on the CCD, because the scan
rate cannot be optimum for all columns of the CCD. Indeed if the
scanning of the CCD and mount are purposefully not synchronized, for example
in order to mitigate saturation of very bright stars by elongating them,
the PSF elongates in the center of the CCD and becomes shaped like a
``)'' or a ``(''
on either side of center. With proper synchronization,
the elongation is nearly imperceptible and nearly uniform across the field
as intended (see Figure \ref{fig:psf}).

%----------------
%-----------

By rotating the mount
at 478\arcsec\ s$^{-1}$about the declination axis while tracking at the
sidereal rate in right ascension, the XO system scans repeatedly from
0\arcdeg\ to 63\arcdeg\ of declination.

The repeatability of a star's position in rows, Y, is directly
related to the repeatability of the timing of the CCD readout and
its synchronization with the equatorial mount. Typically the repeatability is
$\pm$1 second, corresponding $\pm$19 rows, peak to valley.
Occasionally larger excursions occur,
e.g. due to missing the daily reset of one of the computers' clocks. The clocks
on the computers reading the CCDs drift reliably by 4 seconds per day. 

The repeatability of a star's position in columns, X, is directly
related to the repeatability of the mount's positioning in RA. The 
repeatability measured over an observing season is $\pm 4$ columns
($\pm 100$\arcsec), peak
to valley, dominated by a slowly-varying function
of hour angle presumably due to residual misalignments or imperfections
in the mount.

We constructed a ``flat field'' vector from a robust averaging of
measurements of ``sky'' along columns of a few images selected to be
relatively free of stars, saturated stars, Galactic cirrus, and gradients
or curvature. The ``flat field'' for each camera is approximately
parabolic, with the peak within 6 columns of the center of the CCD,
and it is almost entirely due to optical vignetting, because the detectors
are intrinsically uniform and are made even more so by the
drift scanning technique.
At the edges, a noticeable departure upward occurs that
we attribute to excess scattered light from the sky near the edges of
the CCD.  We replaced the edges of the measured ``flat field'' vector
with linear extrapolations from the interior. Other than a bad column
in each of the CCDs, we did not measure any repeatable, small-scale structure
in the
observed flat field vector, so we smoothed it with a 31-column boxcar.
The flat field's peak, near the center column of the CCD, is greater than its minimum
at one of the edges, by 34\% and 35\% respectively for the two cameras.
The absolute value of the slope of the flat field
is everywhere $< 0.9$ millimag~column$^{-1}$.
The slope combined with the repeatability in X-position implies 
repeatability of the instrument of $\pm 3.6$ millimag, peak to valley,
prior to calibration.

\subsection{Mount Control}
\label{subsec:mount}

The equatorial mount selected is a Paramount ME manufactured
by Software Bisque, Inc.\footnote{We experimented with and rejected a
less expensive mount due to its
``runaway'' behavior described by Lopez-Morales \& Clemens (2004).}
We control the Paramount using a custom visual basic script interacting
with {\it The Sky}
by the same manufacturer. The script commands the mount to scan every ten
minutes according
to a pre-determined nightly schedule. For simplicity, the script and the mount
operate regardless of the roof's state.

The Paramount has required no physical maintenance, and
we have experienced only a few minor problems with it.
The cameras are re-oriented by 180\arcdeg\ on
the sky whenever the German equatorial mount crosses the meridian. 
This requires scanning northward east of the meridian and southward west
of the meridian; it also requires calibrating the data separately,
because the positions of a given star on the CCD are not the same on
opposite sides of the meridian.

\subsection{Charge Coupled Devices}

The sensors selected are SITe SI-003AB 1024$^2$-pixel back-illuminated
CCDs with 24 micron pixels, in the model AP8p camera manufactured by
Apogee, Inc. The CCD is cooled thermoelectrically, with waste heat
dissipated by fans on a heat sink.  Because the CCD is illuminated by
a broad optical bandpass with a f/1.8 lens, dark current can be nearly
negligible with moderate cooling, so for simplicity and reliability,
we operate the CCDs in MPP mode at $-30$ C year round.
A disadvantage of the MPP mode for these sensors is that in addition
to ``bleeding'' vertically, bright stars leave trails horizontally.
In order to eliminate the horizontal trails, MPP mode can be suppressed
during readout.  However that option is not suitable for drift scanning
wherein the CCD is read continuously. Operating the CCD without MMP mode
increases the dark current by a factor of 20, which we decided would be
less desirable than having trails associated with a few very bright stars
per image. Each CCD has one unique bad column, but for surveys such as
this, sensors with a single bad column can be more cost effective than
ones with no bad columns.  We flag data for which a star's photometric
aperture contains a bad column.

\section{Observing Strategy and Experience}
\label{sec:observing}

\begin{figure}[!ht]
\begin{center}
\epsscale{1.0}
\plotone{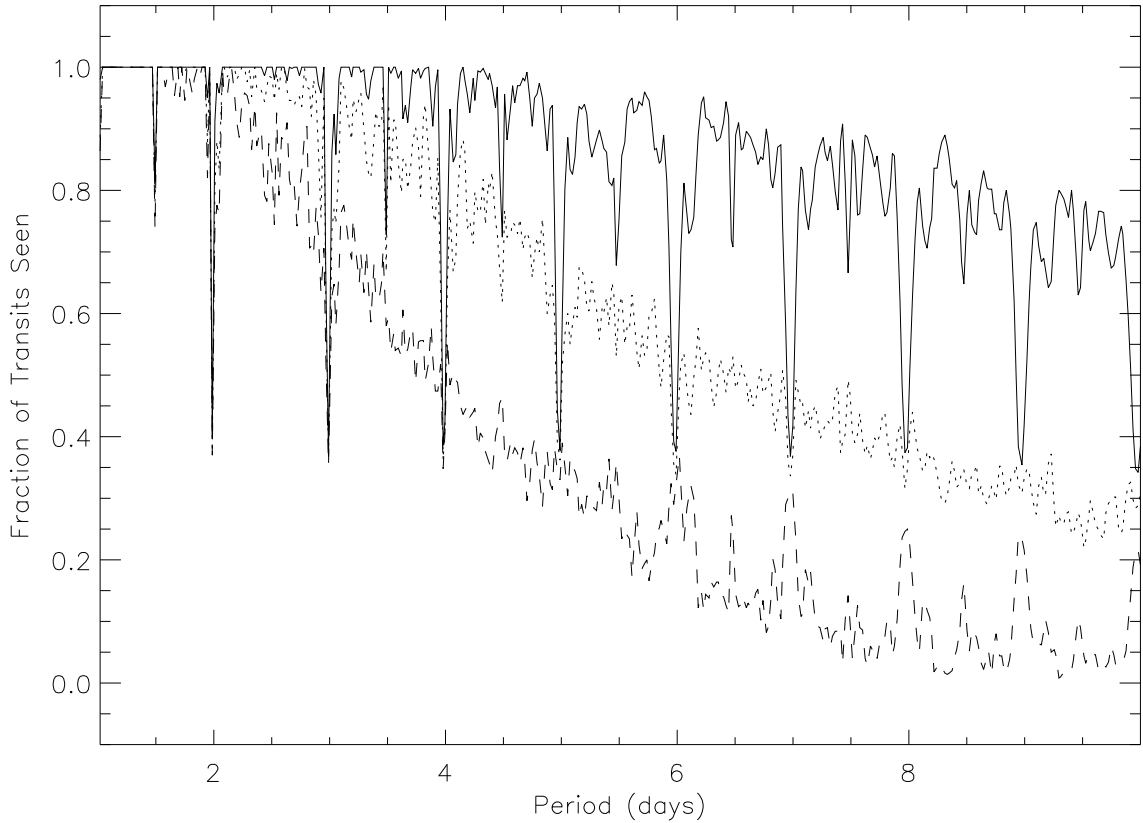}
\caption{The visibility of transits of a given period are indicated for observations of
at least 1 transit (solid),
at least 2 transits (dotted), and at least 3 transits (dashed). In the plot, the fraction of
transits seen indicates the fraction that could be seen with sufficient
detection sensitivity, given the actual times of observation of RA=0$\rm ^h$\ by XO in 2003.
Characteristically, the
visibility is excellent at periods of 3 days or less and tapers off rapidly for longer periods.
\label{fig:visibility}
}
\end{center}
\end{figure}

Our observing strategy is based upon a compromise between observing a region of sky for many weeks and observing at moderate zenith angles. 
As stated before, we scan in declination from 0\arcdeg~to +63\arcdeg, and
each night a program selects the right ascensions $\alpha$ of the scans from a
table.
Each star is observed by each of the two cameras every 10 minutes.
On a given night, we either concentrate on a particular RA, $\alpha_0$, or split the night at midnight and observe one RA, $\alpha_-$, before midnight and another RA, $\alpha_+$, after midnight. 
In the former case, we observe the primary target whenever its hour angle is
within 4 hours of the meridian, and after/before it is within that range we
observe RA=$\alpha_0\pm4$ hours. With this strategy, a season of observing a
given target lasts $\sim$4 months, as follows:
it is observed only a few times in the morning
for the first $\sim$0.5 month; from midnight to morning for the next month;
whenever it is within 4 hours of the meridian for a month;
from evening until midnight for a month; and finally 
only a few times in the evening for the last $\sim$0.5 month of its season.
In a calendar year, this strategy targets six RAs, each separated by 4 hours
from its neighbors. The visibility function as a function of period of the
transits for a representative star is given in Figure \ref{fig:visibility}.
A second season of observation of the same stars improves the visibility of transits,
especially for longer orbital periods, but at the cost of not observing ``new''
stars. With a second observing season on the same stars, the longer time baseline
permits more precise period determination, which is important for predictions of
future transits and for accurately estimating the orbital phase of radial velocities.

%----------------
%-----------

XO's 25\arcsec\ pixel$^{-1}$ scale is approximately twice that of other similar
projects (Figure \ref{fig:image}). This is primarily because our system was designed to scan large
regions of sky, typically far from the Galactic plane, where
the stellar density per unit solid angle is small. Also, the greater
surface brightness in the Galactic plane elevates the photometric noise.
As stated earlier, XO's targets are approximately isotropically distributed
on the sky. Therefore, at low Galactic latitudes we expect that the number of solar type stars to be enhanced somewhat but the number of false positives to
be enhanced greatly, either
astrophysically (Brown 2003) or instrumentally, i.e. that the crowding will make
the photometry more error prone.

In the first year, the XO system's operational readiness was as follows:
1) it missed 36 contiguous nights (10\%)
due to a roof motor failure caused by warping of the roof's rails at the
onset of winter, 2) it did not observe 22 additional nights (6\%) due to 
various shorter-term equipment or software problems throughout the year, 3)
it did not observe at all on 62 nights (17\%) due to weather, and 4)
of the remaining 245
nights (67\%), data was gathered for some fraction of the night, with that
fraction being distributed on [0,1] approximately uniformly.

Figure  \ref{fig:year1} shows many patterns to the actual observations in the first year.
The hour-glass shape is due to the annual pattern of times of darkness for Maui.
The sky is more often cloudy in summer than winter.
The roof was inoperable for all of January 2004.
The telescope stops observing for 10-30 minutes when transitioning from
one target RA to another (which often occurs near midnight)
and at meridian crossing (ragged slanted lines).
Moon light saturates the detector when the telescope scans near the moon,
so some data is lost around full moon; for example, on nights 94-96 moonlight
saturated the southern part of each scan.
The ragged data after midnight on nights 50-55 was caused by a software
error: we had been discarding any image with fewer than 5000 stars detected,
but for RA = 12 hours, this threshold was too high,
so we adjusted it to be 1500 stars.
Our linux operating system performed housekeeping by default at 4 AM,
and until we forced this to occur during daytime, images at 4 AM were
discarded because they were trailed slightly (nights -112 to -89).
During approximately that same period, we had a simplistic schedule
of observing only RA = 0 hours, and we kept the shutter-closed images
taken during twilight.

\section{Software Implementation}
\label{sec:software}

\subsection{Readout of CCD}
\label{subsec:readout}

We accomplish the CCD readout in drift-scan mode using 
Tcl scripts and linux drivers for the Apogee cameras.
We modified the Random Factory Version 0.5 GUI interface
to permit command-line control, because the latter is better suited to
autonomous operation.
We used {\it Xvfb} to create a virtual frame buffer to work around
the code's default operation within X windows.

Each CCD is attached to a laptop computer's parallel port for control
and data transfer. The laptop operates the CCD on a predetermined
schedule, initiated by linux's {\it cron} every 10 minutes.
During twilight
the camera's shutters are not opened but the CCD is readout just as it
will be during normal (shutter-open) operation.
Whenever the sun's
altitude $< -15$\arcdeg, the control computer checks the weather
(Section \ref{subsec:roof}) and if it is clear, it opens the roof
and the camera's shutter and reads the CCD 
at a rate of 18.8 rows per second. The first 1024 rows are discarded
because they are improperly exposed, and the next 9000 rows are
stored as a FITS file. These files are automatically analyzed the following
morning as described in the next Section.

Rarely, one of the linux laptops ``freezes'' and must be revived by human
intervention.
On the summit of Haleakala, the laptops are at their specified limit
for ambient atmospheric pressure (altitude $< 3048$ m), and for the
first year they were routinely operated at too high an ambient temperature,
especially for a few hours each morning when they reduced data.
The frequency of ``freezes'' has been reduced to once every few months
by increasing ventilation to reduce the temperature of the laptops' ambient
environment by $\sim$5 C to within specification (T $< 31.3$\arcdeg\ C).

Initially we operated the CCD's only at night, warming them in the morning and
cooling them at night. This was to reduce the risk of a power outage causing
an uncontrolled return to ambient temperature that might crack the CCD or
its thermoelectric cooler.
However, occasionally a CCD would not cool after turning it
on at nightfall, so we chose to leave them on and cooled all the time.
XO has experienced at least one power outage that outlasted the UPS, but
no equipment was damaged.

% ========================== DATA REDUCTION ==============================
\subsection{Data Reduction to Instrumental Magnitudes}
\label{subsec:datared}

Although the data are taken as 1024x9000 pixel arrays, for convenience we
carve them into nine 1024x1024 images for further analysis. We carve
the images at a 1000-row pitch, which allows for 24 pixels of overlap
between images so that we do not loose stars due to the carving. During
twilight we drift-scan images in the same manner as we do during the
night except that we do not open the shutter. The last one exposed
each night is used to create a ``dark'' vector to subtract from all the
images taken during the night.  We constructed a ``flat field'' vector
as described in Section \ref{sec:hardware}.

For each dark-subtracted and flat-fielded image, we use
stars to determine the astrometric solution and add it to the header
(Gaustad et al. 2001). Also we add keywords related to
the weather and the positions of the Sun and Moon. Next, we use
{\it Stellar Photometry Software} (SPS, Janes \& Heasley 1993)
to find stars by searching for peaks in the image's cross
correlation with a nominal Gaussian PSF. For each such star SPS measures
the centroid in X and Y on the CCD, the instrumental magnitude using
aperture photometry, the estimated error in that magnitude, the local
sky level and the correlation coefficient of the star's PSF with the
Gaussian PSF used to find the stars.

A representative line of data stored for a given observation of one star
is in fixed ASCII format, ``1008.73 1001.02 12.7234 0.0028  2280.85''
for the X, Y, Mag, Error, Sky+CC, respectively. The last quantity encodes
the sum of the sky level rounded to the nearest integer and the
correlation coefficient. Using this simple ASCII format, each star's
photometry requires 40 bytes uncompressed. Using standard gzip compression,
the file compresses by a factor X, where X = 2.12+0.3log$_{10}$S, where S is
the size of the ASCII file in kilobytes. For typical file sizes of 250 kB,
the compression factor X = 2.8, so each star requires 14 bytes of information
per camera per epoch. 

Each day the previous night's photometric measurements from SPS and the
associated FITS headers for each image are transmitted to STScI
for further calibration, which is described in the next section. All
of the calibrated pixel data and the photometry of all of the stars
detected in each image are stored on external disks that are mailed
to STScI when they are filled, typically every 3 months.
The conversion from raw to calibrated pixels is reversible to
$\pm$1 ADU accuracy, so we discard the raw data and store the calibrated
pixel data after applying lossless compression with ``encode'' (Sabbey 1999).

In the first year of operation, we gathered $\sim$132,000
1-Mpixel images that
were of sufficient quality for our pipeline to recognize a star pattern.
The data volume after lossless compression is 184 GB.
During a clear, 10-hour, winter night the system gathers 1100 1-Mpixel images.
Transmitting all the pixel data via the internet is possible
but would use a significant fraction of the bandwidth from the summit, so the
data-taking computers reduce the data from pixels to stellar photometry each
morning and only transmit the photometry, which is 15 times fewer bytes
than the pixels.

\subsection{Data Organization and Visualization}

The first part of the calibration pipeline arranges the
SPS output from each image and its associated FITS header into an organized format.
For each field we hand-pick a single image taken on a clear moonless night and
extract a list of reference stars. 
The reference list defines the stars that we study for that
field. Very many stars observed and analyzed with SPS
are ignored by our pipeline because they are too faint to be on the list of
reference stars.
In our current implementation,
we select stars with instrumental magnitude
less than 16 (corresponding to V $<$ 13.3), and also we
select only the brightest 5000 stars in each field. 

The stellar photometry from each night is arranged into 2-D arrays stored
as FITS files, with one axis indexing stars and the the other axis indexing
epoch. 
Each photometric attribute that is appropriately stored for each
star and epoch is stored as a separate FITS file. The attributes
stored in this manner are X position, Y position, instrumental magnitude, predicted error in
the magnitude, sky brightness, cross correlation coefficient, RA, DEC, and airmass.
By ordering the stars according to their apparent magnitude, and ordering
the epochs in time, we can conveniently see trends or errors in calibration
by visualizing the photometry as an image (Figure \ref{fig:starvsepoch}).

An accompanying 2-D FITS file stores information associated with each image and
epoch as a whole, extracted from the FITS header. Example keywords are
those describing the astrometric solution,
the Julian date, the weather (humidity, temperature,
pressure, etc), the positions of the Sun and the Moon,
a prediction of the sky's surface brightness due
to moonlight (Krisciunas \& Schaefer 1991), etc.
In total $\sim$60 such keywords exist for each image.

\subsection{Calibration}

The calibration is based upon these principles: 1) the drift-scan observing method 
inherently should provide repeatable and precise measurements, 2) only
differential magnitudes are required, 3) at the amplitudes and on the timescales
in which we are interested in, the majority of stars are constant in brightness and
can be used as calibrators, and 4) each star will have its own unique set of
comparison stars for differential, ensemble photometry.

The calibration is done iteratively. First we apply nominal calibrations to
all stars, such as nominal airmass corrections, low-spatial-frequency corrections
to the flat field, and offsets to bring the two cameras' magnitudes into agreement.
The sole purpose of this ``zeroth order'' calibration is to permit us to assign
comparison stars that are nearest to the target
star in brightness and in position, i.e. in columns of the CCD. The justification
for this method is that stars of similar brightness that traverse the CCD
at nearly the same location should make good comparison stars because uncorrected
errors of calibration will be made equally to them all and
will vanish in differential measurements. Because
brightness and position cannot be directly compared, we determine the
``nearest'' stars by repeatedly bifurcating the sample of stars, first
by distance in columns, then by difference in magnitude, and at each
bifurcation, retaining the nearer half. We repeat the process as long
as the sample exceeds 80 stars, from which we pick the 10 nearest to
the target star's magnitude as comparison stars.

Now that we have assigned comparison stars, we begin again the calibration
starting with the raw instrumental magnitudes.
Because each CCD has its own instrumental signature and because the 
CCD's orientation on the sky changes by 180 degrees on opposite sides of the
meridian due to the German equatorial mount, 
we perform the calibration for four cases distinctly: 1) CCD 0 scanning south,
2) CCD 1 scanning south, 3) CCD 0 scanning north, and 4) CCD 1 scanning north.
Near the end of the calibration we combine the four cases.
We apply a nominal correction of 0.086 magnitudes per airmass.
Next, an empirically-determined third-order correction to the flat field is applied. Equivalently we could
have adjusted the flat field vector itself, but that would necessitate
re-processing all the pixels again, so it is more convenient to adjust the magnitudes.
We subtract from each and every star's light curve its average magnitude,
so hereafter each star's average magnitude is zero, facilitating comparisons.
(Here and elsewhere, ``average'' refers to a resistant
mean with outlier rejection.) At each epoch we subtract from the target star's
magnitude the average magnitude of its comparison stars for that epoch.
We use a robust average as described previously, but a possible enhancement
would be to use a weighted
average as described by Kov{\' a}cs et al. (2005).
Also for each epoch we use stars with ${\rm 10 < V < 11}$ to determine a
third-order polynomial fit
to the residual magnitudes with respect to the column on the CCD, and we subtract
that fit from all the magnitudes at that epoch.
For each star, we fit and subtract any linear dependence on airmass that
remains; we presume this corrects for color
differences between the target star and its comparison stars.
For XO's 0.4 $\mu$m to 0.7 $\mu$m bandpass, maximum
zenith angle of $\sim$60 degrees, maximum angular separation between
target star and reference star(s) of 7\arcdeg, the expected residuals due to
differential airmass between target and reference are small, $\la 2$ millimag.
Because we choose our reference stars so that they'll have similar CCD
columns, they tend to have similar right ascensions, and in our case that tends
to make the differential zenith angle much smaller than the angular
separation of the target from the reference(s) at large hour angles where
the correction is most important.  
Figure \ref{fig:starvsepoch} illustrates the data
after calibration and before flagging.

We flag entire epochs for which 
the rms residuals of the stars with ${\rm 10 < V < 11}$ exceeds 12 millimag.
We flag particular data points for which the SPS-estimated error
exceeds 15 millimag,
or for which the sky brightness exceeds 10000 ADU.
The flagging eliminates most of the
data for nights within a couple days of full moon, but after the period
searching is finished (Section \ref{subsec:searching}), these points can be
re-examined by a human who can
more effectively ignore the uncalibrated effects of the moonlight.
We selected the red edge of our spectral bandpass (0.4 $\mu$m to 0.7 $\mu$m)
in order to avoid the rapid increase of the brightness of the night sky
in the near infrared and the telluric absorption bands therein. 
For a simple filter that transmits light shortward of a
wavelength $W$ in microns, while holding all other characteristics of our
system constant and optimizing for the spectrum of the night sky with no
moonlight, we predict optimum sensitivity to V=12 solar-type stars with
$W = 0.77$, with values of $W$ between 0.7 and 0.85 giving signal to noise
ratios within 10\% of optimum. To reduce saturation from moonlight
and to avoid potential variability of the telluric absorption bands,
we selected $W = 0.7 \mu$m.

In Figure \ref{fig:starvsepoch} data are missing due
to clouds (horizontal lines), saturation (e.g. ragged left edge near center),
a star positioned near a bad column in one CCD but not the other (vertical
``dotted'' line segments), or the star's reference position being erroneous
(vertical lines). Prior to calibration, trends are evident such as 
extinction varying with airmass (nightly undulations in the grey scale) or
noise increasing
with sky brightness (monthly cycle due to the moon). The effects of 
calibration and thresholds for flagging can be visualized with images
of this type. For example, one can see that after our nominal calibration, 
the brightest stars (first
$\sim$100 columns from the left) still have more noise than 
expected, because the grey scale is smoother for somewhat fainter stars
(columns 100-300). Presumably the cause is that the peaks of the brighter stars
saturate the CCD, especially when the moon elevates the sky level on the
CCD.

\subsection{Searching for Transit-like Light Curves}
\label{subsec:searching}

We use a modified form of the Box Least Squares (BLS) algorithm to search
light curves for transit-like signatures (Kov{\' a}cs et al. 2002).
The core of the algorithm is the FORTRAN subroutine ``bls.f'' modified by one of us (P. R. M.) to correct a small error in the original that did not treat
transits near phase zero properly. Kov{\' a}cs renamed it ``blsee.f,''
tested it, found it to execute approximately as fast as the original, and
distributes it on the internet alongside the original. 

We call the BLS code from within IDL,
thereby allowing the flexibility of IDL with the speed of FORTRAN. 
Because the execution time for the BLS routine is linear with the number of
frequencies searched, and because generally we wish to measure only the peak(s)
of the BLS spectrum, we have created an adaptive mesh search routine that
first creates a
nominal BLS spectrum using 5400 frequencies between 0.1 day$^{-1}$ and
0.95 day$^{-1}$, with spacing ${\Delta}f$ such that $f/{\Delta}f = 2400$.
We chose the frequency resolution
$f/{\Delta}f = 2400$ in order to resolve peaks in the BLS spectrum created
by two dips in the light curve, each two-hours long, separated by 100 days,
which is approximately
the duration of an observing season for one field of view.
The adaptive mesh searches near the 40 highest local maxima of the
spectrum in order to quickly find the highest of those.
While the 0.1 day$^{-1}$ may be justifiable based upon the diminishing
sensitivity of our observing window to longer periods, the 0.95 day$^{-1}$ limit
is a error of expediency that we intend to correct in future analyses by extending the search to frequencies greater than 1 day$^{-1}$ and using a notch filter to ingore frequencies very near 1.0 day$^{-1}$.

\section{Verification}
\label{sec:ver}

\begin{figure}[!ht]
\begin{center}
\epsscale{1.0}
\plotone{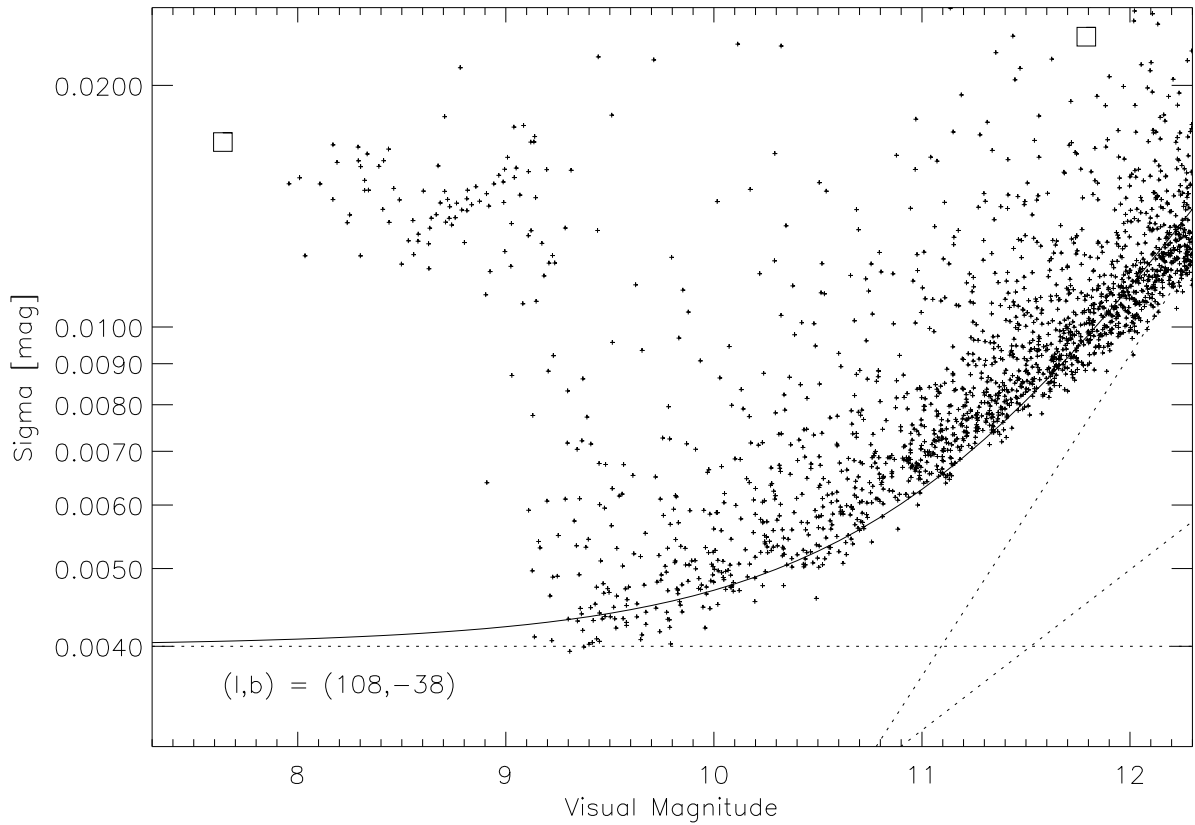}
\caption{The standard deviation $\sigma$ of the differential instrumental
magnitudes estimated by each camera in a 54-second exposure made every
10 minutes is plotted with respect
to the visual magnitude of each of the 1874 stars in a representative field
at moderate Galactic latitude. Galactic coordinates of the center of the
field are indicated. The $\sigma$ is estimated from
$>$1000 observations per star, after calibration and
iterative rejection of outliers.
The expected performance (e.g. Everett \& Howell 2001)
is plotted as a solid line, which is the quadrature sum of three components:
Poisson noise of photons from the sky (steeper sloped line) and the star (shallower sloped line) and a constant fractional error (horizontal line) attributed to
scintillation and imperfect calibration.
Stars brighter than V$\approx$9 are saturated and have
$\sigma \approx 0.015$ mag.
Two stars known to exhibit transits, HD 209458 and TrES-1, are illustrated
as squares located at their mean brightnesses and transit depths in V band.
\label{fig:emagvsmag}
}
\end{center}
\end{figure}

Figure \ref{fig:emagvsmag} demonstrates that the XO system
meets the photometric requirement described in Section \ref{sec:rqmt}.
The most important component of noise is the Poisson noise due to the
sky brightness integrated within a 3-pixel (76\arcsec) radius of the
photometry aperture. The radius was selected as a compromise between
smaller apertures optimized for fainter stars and larger apertures optimized
for brighter stars. For simplicity we did not adjust the aperture according to
the brightness of the star, and because the PSF is spatially dependent and
occasionally can be slightly elongated due to imprecise scanning of
the telescope mount, we selected the best single-radius compromise.
For those same reasons we did not attempt to implement image-subtraction
techniques (e.g. Alard 2000). 
For the XO observations, scintillation is expected to be $\la 0.003$ mag
(Dravins et al. 1998, Eq. 10).
In Figure \ref{fig:emagvsmag}, the two slanted
lines, derived from XO observations calibrated absolutely to LONEOS catalog
stars (Skiff \& Richmond 2003), cross at V=10.7,
which is the magnitude of a star whose light equals
that of the night sky in the r=76\arcsec aperture. This corresponds to
a sky brightness of V=21.4 per \sqarcsec, which equals the
value independently measured for the dark sky near the zenith observed from
2800-m level on Mauna Kea at the same phase in the solar activity cycle
(Krisciunas 1997).
One penalty of the large photometric aperture
is that during times of bright moonlight, the sky brightness can degrade
the photometry considerably. For lunar phase within 30\arcdeg\ (i.e. 2.5 days) 
of full, the moonlit sky is brighter than the dark sky by 2 mag or more
(Krisciunas \& Schaefer 1991) and the XO observations are not useful
for planet transit observations, as can be appreciated by considering the
effect of shifting the steeply-sloped line in Figure \ref{fig:emagvsmag}
to the left by 2 mag or more.

In Figure \ref{fig:emagvsmag}, the smallest values of $\sigma$ in the
field are $\sim$0.0045 mag; the equivalent values for fields near the
Galactic pole or plane are $\sim$30\% lower or higher, respectively.
Presumably stellar crowding is adversely affecting the aperture
photometry, especially at low Galactic latitudes.
The increased ``sky'' brightness due to the Galaxy increases $\sigma$
by as much as a factor of two from that plotted in Figure \ref{fig:emagvsmag}
for stars with $|b| \la 10$ and V$>$11. The latter effect
is offset by the variation of the number density of stars as with $|b|$ such 
the {\it number} of stars for which $\sigma < 0.014$ mag peaks at
$\sim$60 stars per \sqarcdeg\ at $|b| \approx 20$\arcdeg\ and is
$\sim$20 and $\sim$30 stars per \sqarcdeg, respectively near the Galactic
poles and plane. While those values are empirically determined for
the XO system, the functional dependence of $\sigma$ on the dominant
components of noise (see Figure \ref{fig:emagvsmag}) is sufficiently good
that one could optimize a observing system for a given range of $|b|$.
While XO was designed for moderate $|b|$, other systems have been
optimized for $|b| \approx 0$\arcdeg\ by using smaller pixels (in arcseconds)
and demanding a sharper PSF and thereby allowing smaller synthetic
apertures for their CCD photometry. 

In Figure \ref{fig:emagvsmag}, the sparse envelope above the predicted noise curve
includes stars with elevated noise, in addition to stars with
instrinsic variability.
Several factors cause the elevated noise, including bleeding of charge
from bright stars, blending of stellar images, vignetting, etc.
We have not yet thoroughly investigated all of these
effects because even with the elevated noise, we have adequate
sensitivity to detect planetary transits from a large sample of stars.
Light curve analysis then eliminates candidates with sources
of variability other than transits. Prompt THJ discovery is important,
so that the Spitzer and Hubble Space Telescopes are still available
for precise followup observations.

\begin{figure}[!ht]
\begin{center}
\epsscale{1.0}
\plotone{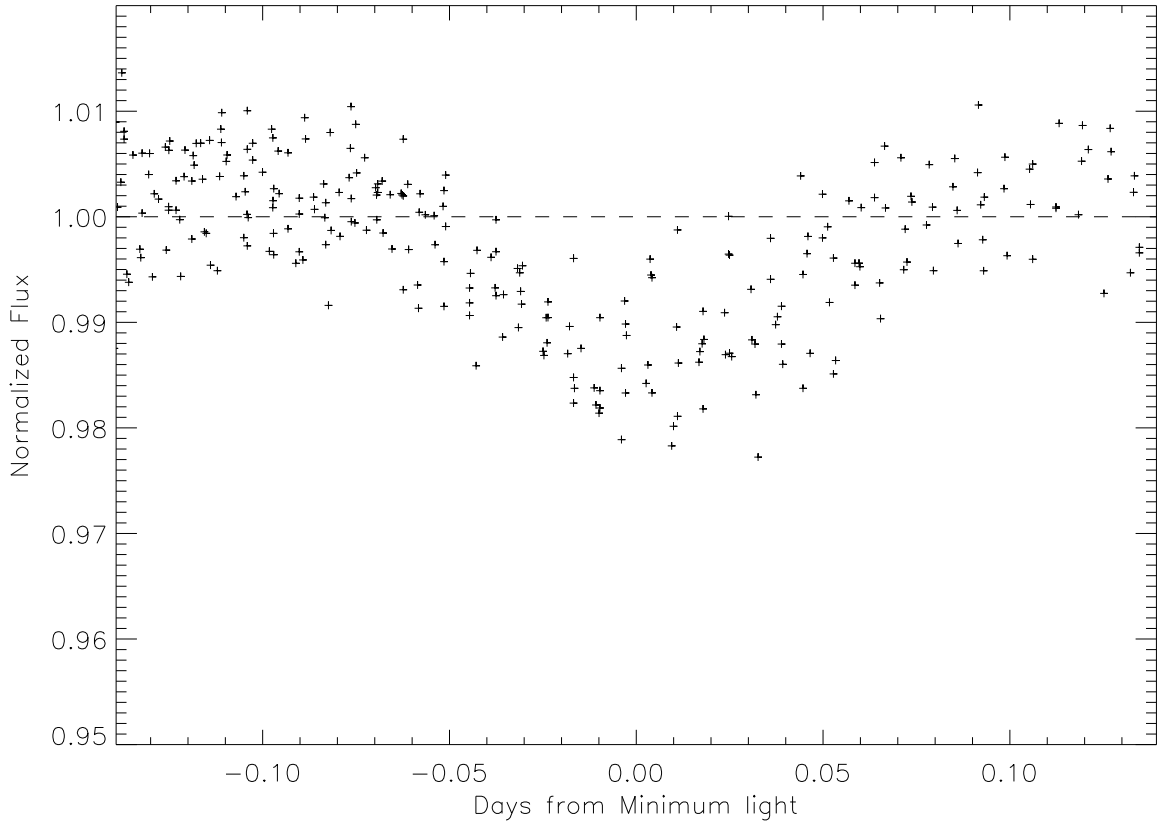}
\caption{XO's calibrated light curve of a 10th magnitude star
exhibits 1.5\% dips for
$\sim$2 hours with a period of 2.79 days. Each symbol represents a single observation from one XO camera. Observations from both cameras are plotted.
\label{fig:lightcurve}
}
\end{center}
\end{figure}

\begin{figure}[!ht]
\begin{center}
\epsscale{0.8}
%\plotone{fg10.ps}
\includegraphics[scale=.50,angle=90]{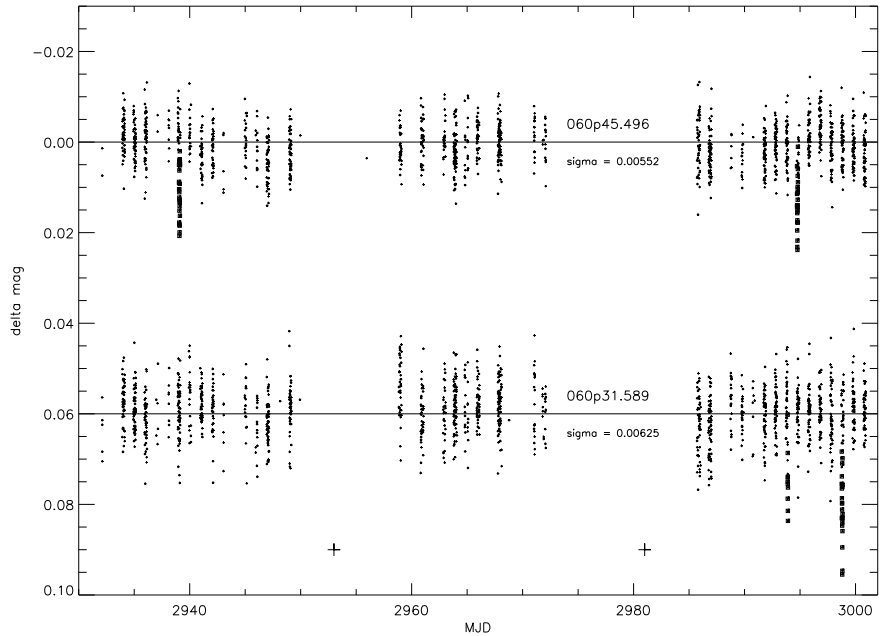}
\caption{Light curves are shown for two stars as functions of time in days.
Observations with phases corresponding to the transits are indicated with open squares.
The lower curve has been shifted down by 0.06 mag.
These light curves are shown binned and folded in the next figure.
The label 060p45.496 encodes the approximate RA, DEC, both in degrees, followed
by an integer that counts the star-like objects from brightest to faintest in
that field of view.
\label{fig:jdhlightcurves}
}
\end{center}
\end{figure}

\begin{figure}[!ht]
\begin{center}
\epsscale{0.8}
%\plotone{fg11.ps}
\includegraphics[scale=.50,angle=90]{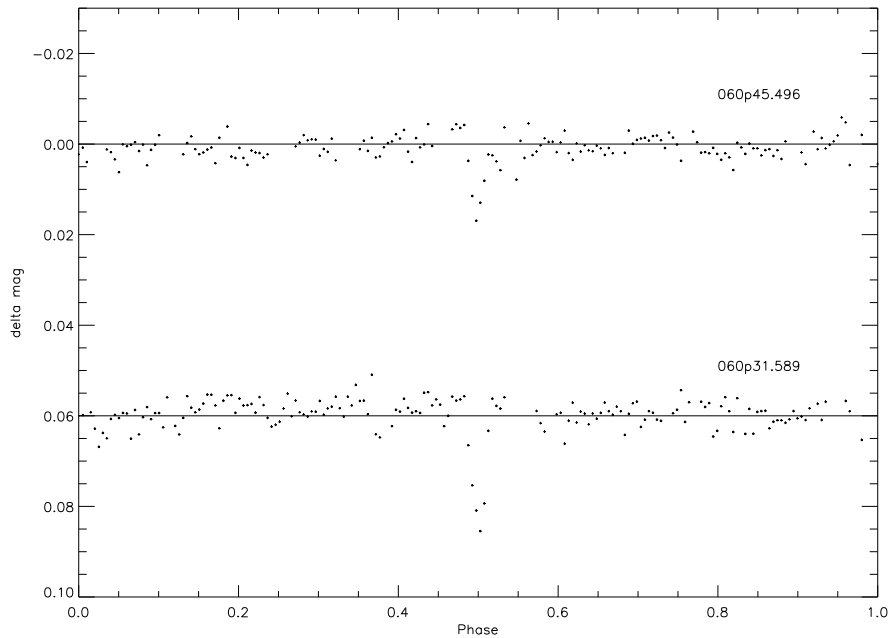}
\caption{Light curves folded with appropriate periods are shown as functions of orbital phase,
shifted such that the dips are centered at 0.5. The data have been averaged into
200 uniformly-spaced bins for clarity. The lower curve has been shifted down by 0.06 mag.
These light curves are shown unfolded in the previous figure. A portion of the upper light
curve is shown without binning in \protect{Figure \ref{fig:lightcurve}}.
\label{fig:twolightcurves}
}
\end{center}
\end{figure}

\begin{figure}[!ht]
\begin{center}
\epsscale{1.0}
\plotone{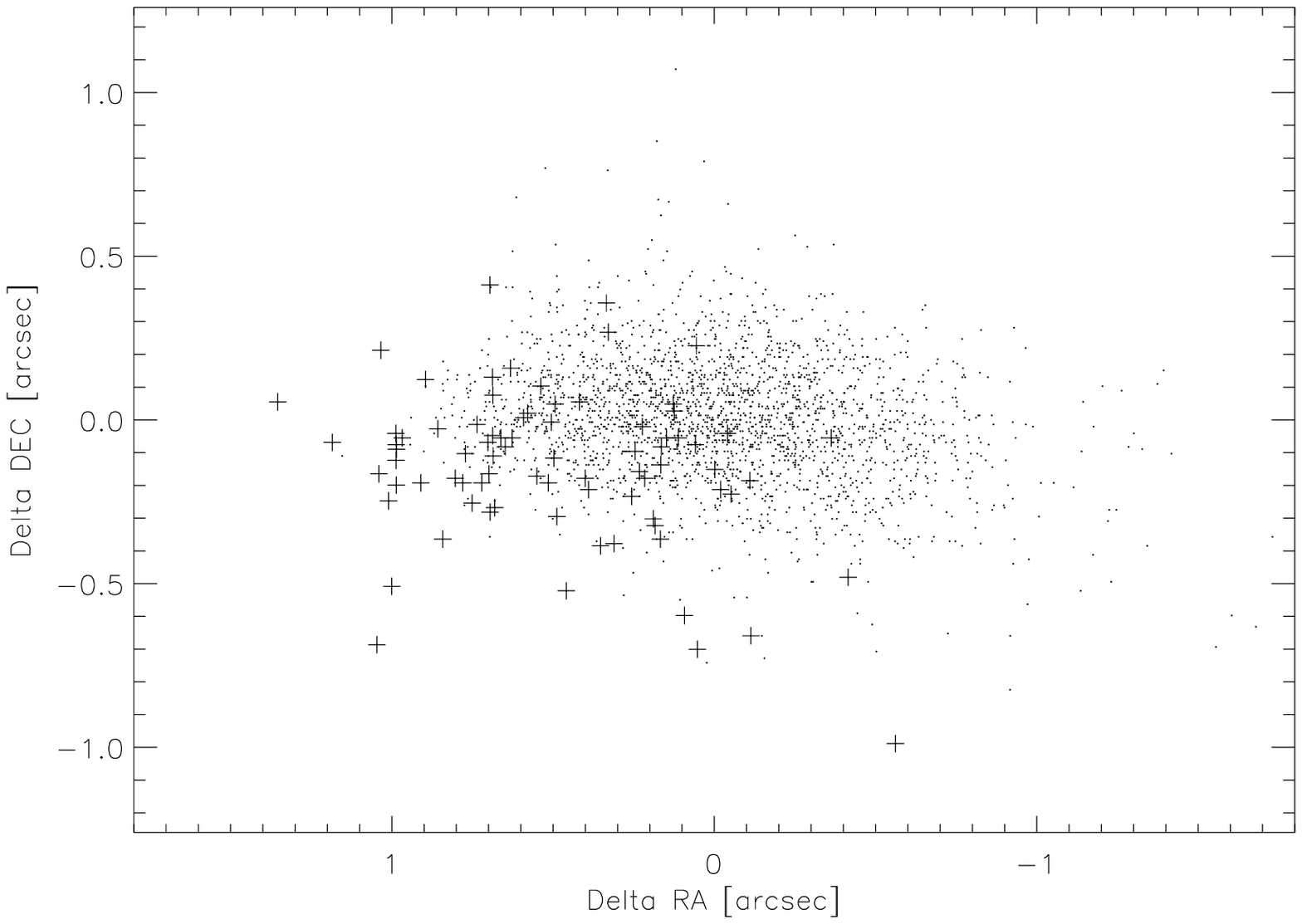}
\caption{The astrometric position of a star observed with XO is
repeatable to $\sim$0.25\arcsec\ (0.01 pixel) rms. Here each measurement
is plotted, and those observed during the periodic dips in the light curve are
indicated with $+$ symbols. That the distribution
of $+$ symbols is shifted significantly to the lower left (southeast) of
the rest of the measurements suggests that a stellar blend is responsible.
Digitized Palomar Observatory Sky Survey plates and 2MASS images
show a star 42\arcsec\ to the northwest and 0.6 mag fainter that could be
responsible:
a self-consistent model is that the fainter star's flux drops by 5\% which
induces both the observed 2\% drop in the combined flux and
the observed shift of 0.5\arcsec\ in the centroid of the pair's blended light.
In cases such as this one, definitive identification of the variable star
is straight forward with a telescope with $\sim$1\arcsec\ resolution.
\label{fig:astrom}
}
\end{center}
\end{figure}

Figures \ref{fig:lightcurve}, \ref{fig:jdhlightcurves},
and \ref{fig:twolightcurves} show light curves for two of many
XO transit candidates. The light curve in Figure \ref{fig:lightcurve} has the
depth, but not the shape, expected of a THJ. Nonetheless, the figure
demonstrates that the XO system has the sensitivity and sampling required
to detect planetary transits. As XO's temporal and angular coverage
continues to increase, the odds of discovering one of the $\sim$8 THJ
systems with $V \le 10$ (Section 2) increases. A fundamental challenge
is distinguishing actual planetary transits from a large number of false
positives caused by a variety of non-planetary transits (e.g., Seager \&
G. Mallen-Orn\'elas 2003; Torres et al.\ 2004; 2005). The light curve
shown in Figure \ref{fig:lightcurve} is more V-shaped than U-shaped,
suggesting a grazing eclipse in a binary system or an eclipsing binary
diluted by a third star. In fact, higher resolution images reveal a
visual binary with a separation of $\sim 1$\arcsec, which is too small
to induce a shift in the centroid measurable with XO as illustrated in
Figure \ref{fig:astrom}.  McCullough et al. (2005) describe in detail
our followup strategy and results for this and other transit candidates
detected by XO.

The astrometric repeatability of well-exposed and reasonably
isolated stars in the XO images is $\sim$0.25\arcsec\ (0.01 pixel) rms.
Due to distortion in
the lenses and differential refraction in the atmosphere, a third-order
polynomial correction to the nominal linear plate equations
is required to achieve this precision. A shift in the astrometric position
in phase with the dip in the light curve may indicate that 
a nearby eclipsing binary star is causing both the shift and the dip
(Figure \ref{fig:astrom}).

%----------------
%-----------

%----------------
%-----------

%----------------
%-----------

%----------------
%-----------

%----------------
%-----------

%% ============================= SUMMARY =================================
\section{Summary}
\label{sec:sum}

A system of two small photometric cameras on a common equatorial has been constructed and operated autonomously on Haleakala, HI, for more than one year. The system is designed to find hot Jupiters transiting bright
stars, i.e. V $<~12$, by precision differential photometry.
Some unique aspects of it are its two identical cameras, each with
a broad spectral bandpass (0.4 $\mu$m to 0.7 $\mu$m), mounted together in 
binocular fashion on a mount that
scans at 478\arcsec\ s$^{-1}$ to cover large areas of the
sky and to simplify calibration.
We have searched for the periodic dimming of transiting exoplanets
in the light curves of approximately one third of
the $\sim$100,000 stars observed with photometric precision per observation
from 0.004 mag to 0.015 mag rms for stars of visual magnitudes
9 to 12 respectively. 
Analysis of additional photometry and spectroscopy of the resulting exoplanet
candidates is described in McCullough et al. (2005).

% ======================== ACKNOWLEDGEMENTS ==============================
\acknowledgements

A large number of persons have made significant contributions to the XO project.
Beth Bye and Chris Dodd helped deploy the initial prototype on Maui.
The University of Hawaii staff have made the operation on Maui possible; we
thank especially Bill Giebink, Les Hieda, Jeff Kuhn, Haosheng Lin, Mike Maberry, Joey
Perreira, Kaila Rhoden, and the director of the IFA, Rolf-Peter Kudritzki.
We benefited from discussions with Gaspar Bakos, Ron Bissinger, Fred Chromey,
Bruce Gary, Ron Gilliland, Leslie Hebb, James McCullough, Margaret Meixner,
Kailash Sahu, and Tonny Vanmunster.
We benefited from the following computer programs that are acknowledged because
in addition to being essential, they are also freely distributed on the
internet:
{\it rsync} by Andrew Tridgell, Paul Mackerras, et al.,
{\it vnc} by Tristan Richardson, et al. (1998),
{\it ssh 2} by Markus Friedl, et al.,
transit light-curve simulators by Mandel \& Agol (2002),
{\it bls} by Kov{\' a}cs et al. (2002),
and
{\it match} by Michael Richmond.
 
XO is funded primarily by the Origins program of NASA
(NAG5-13130); in the past decade, funds have been provided by
the Sloan Foundation, the Research Corporation, and the US National Science Foundation.

% ============================ BIBLIO ====================================

\end{document}